\documentclass[copyright,creativecommons]{eptcs}
\providecommand{\event}{QAPL 2019} 
\usepackage{soul}
\usepackage{comment}

\title{Quantitative Aspects of Programming Languages and Systems over the past $2^4$ years and beyond}
\author{Alessandro Aldini
\institute{University of Urbino\\ Urbino, Italy}
\email{alessandro.aldini@uniurb.it}
}
\def\titlerunning{QAPL over the past $2^4$ years and beyond}
\def\authorrunning{A. Aldini}
\begin{document}
\maketitle

\begin{abstract} 

Quantitative aspects of computation are related to the use of both physical and
mathematical quantities, including time, performance metrics, probability, and
measures for reliability and security. They are essential in characterizing the
behaviour of many critical systems and in estimating their properties. Hence,
they need to be integrated both at the level of system modeling and within the
verification methodologies and tools. Along the last two decades a variety of
theoretical achievements and automated techniques have contributed to make
quantitative modeling and verification mainstream in the research community. In
the same period, they represented the central theme of the series of workshops
entitled Quantitative Aspects of Programming Languages and Systems (QAPL) and
born in 2001. The aim of this survey is to revisit such achievements and
results from the standpoint of QAPL and its community.

\end{abstract}

\section{Introduction}

Quantitative aspects of computation refer to the use of physical quantities,
like, e.g., time and bandwidth, as well as mathematical quantities, like, e.g.,
probabilities, for the characterisation of the behaviour and for determining
the properties of systems. These quantities contribute to define both the model
of systems from different perspectives (architecture, language design,
semantics) and the methodologies and tools for the analysis and verification of
system properties. 
In the last two decades several tool-supported developments, novel frameworks, 
and theoretical results in this area have been achieved. 
This is also witnessed by the amount of domains, ranging from security to biology 
and from quantum computing to hardware analysis, to which quantitative models and 
formal verification approaches have been applied.

The objective of this survey is to provide an overview of such achievements,
their effects on the research trends, and the open issues they have left,
through the lens of QAPL, the International Workshop on Quantitative Aspects of
Programming Languages and Systems, born in 2001 as one of the first events in
theoretical computer science to focus on the design of models and techniques
capturing the various quantitative aspects of computation. Nowadays, the
scientific research on such quantitative aspects is mainstream for a majority
of communities and leading conferences on the foundations of programming
languages and systems.

Through the analysis and classification of the contributions presented at QAPL,
we revisit some research topics concerned with the quantitative modeling
and verification of programming languages and systems, by offering a survey with
respect to the following categories: languages and models, logics, information 
flow analysis and equivalence checking, model checking, static analysis, and 
related verification tools. 

\section{Trends}

The early 2000s, which cover the period in which the proceedings of the
workshops have been published as volumes in Electronic Notes in Theoretical
Computer Science (ENTCS), concentrate on several formal frameworks for the
quantitative modeling and analysis of systems, the main representatives of
which can be categorized as follows: 
concurrent constraint programming (6 papers), process algebraic languages 
(12 papers), modal logics (11 papers) and model checking (6 papers), 
static analysis based on abstract interpretation (6 papers), 
behavioural equivalences and their approximations (8 papers), 
information flow analysis (9 papers). 
Classical techniques, like numerical analysis and performance evaluation, 
received attention from time to time.
During this period, 72 papers were published, collecting 924
citations, CiteScore equal to 12.83 and
h-index equal to 15\footnote{Source: Scopus in May 2019.}.

In the second decade of 2000s, QAPL proceedings appeared as volumes of the
Electronic Proceedings in Theoretical Computer Science (EPTCS). By respecting
the same classification emerged above, the research on process algebra (8 papers)
and other enhanced formal paradigms (9 papers), behavioural equivalences and
information flow analysis (15 papers), has continued its upward trend. Topics
like concurrent constraint programming almost disappeared, while it is worth
observing that the focus moved significantly towards the study and development 
of a variety of efficient analysis techniques, supported by automated tools, 
in the setting of static program analysis and abstractions (8 papers), model 
checking (9 papers), game theory (7 papers), numerical and approximate analysis
of Markovian and hybrid systems (8 papers). 
During this period, 58 published papers produced 261 citations, corresponding to 
CiteScore equal to 4.5 and h-index equal to 9\footnote{The 2011 proceedings 
collecting 11 papers are not indexed and, therefore, not included in the statistics. 
Source: Scopus in May 2019.}.

\subsection{Languages and Models}

Expressiveness of modeling languages is an important and intriguing issue going
beyond the pure functional aspects of sequential and concurrent
programming. In many settings, the purpose is not only to model the
functionalities of a system, but also to quantify its behaviour in order, e.g.,
to enable performance analysis. 

By enhancing programming language semantics with probabilities, the properties
of programs are not ensured with certainty but up to some probability.
A commonly used framework for such a kind of extension is given by the 
probabilistic guarded command language pGCL and its related quantitative
logic. In such a setting, properties are expressed through a logic of real-valued
functions, called expectations, and several techniques are used to verify them
efficiently~\cite{DBLP:journals/tcs/HurdMM05}, including abstractions~\cite{EPTCS28.9}.

Concurrent constraint logic programming is a paradigm for reasoning about
concurrency tied to logic and for computing with
constraints~\cite{Saraswat:1989:CCP:96709.96733}, which represent partial
information about variables that describe every state of the program execution.
In the setting of constraint logic programming language, a general complexity
meta-theorem for the worst-case time complexity of programs is
known~\cite{FRUHWIRTH2002185}.  Timed extensions of concurrent constraint
programming language introduce temporal aspects enabling the model checking
analysis of timing properties of concurrent systems (see, e.g.,
\cite{BORTOLUSSI200665}, where both discrete and continuous time are modeled).
As usual in such a kind of extensions, abstract techniques are defined to deal
with the state space explosion problem~\cite{DBLP:journals/tcs/AlpuenteGPV05}
or to enhance the analysis capabilities, e.g. through semantics based on
ordinary differential equations~\cite{BORTOLUSSI200727,BORTOLUSSI2008163}.
Constraint logic programming is also combined with constraint-semiring
structures, which consist of a domain (representing the values related to
constraints), an additive operation (for projecting constraints), and a
multiplicative operation (for combining constraints), so that to enable
modeling of quantitative aspects such as costs~\cite{BISTARELLI2007111}.

A lot of work has been done to endow process algebraic approaches with
quantitative elements describing temporal and/or probabilistic behaviours. 
Such studies go back to the early 1990s and reached their full maturity in the
2000s, where it is easy to find an amount of extensions and variants of both
qualitative and quantitative languages. In the setting of QAPL, we can mention
several such examples: 
probabilistic variant of the process-algebraic $\mu$CRL
language~\cite{KATOEN201236}, discrete time variant of distributed
$\pi$-calculus~\cite{CIOBANU200681}, probabilistic extension of
$\pi$-calculus~\cite{PRADALIER2006119}, interleaving semantics and true
concurrent semantics for the probabilistic variant of
$\pi$-calculus~\cite{VARACCA2007147}, stochastic broadcast
$\pi$-calculus~\cite{EPTCS57.6}, stochastic version of Mobile
Ambient~\cite{VIGLIOTTI2006169}, stochastic extension of the hybrid process
algebra HYPE~\cite{EPTCS57.9,EPTCS85.8,EPTCS154.5}, stochastic extension of the
Software Component Ensemble Language for modeling ensemble based autonomous
systems~\cite{EPTCS154.1}, Linda-like coordination calculus extended with
quantitative information~\cite{DBLP:journals/tcs/BravettiGLZ05}, and finally a
mixture of concurrent and probabilistic Kleene algebras enriched with
probabilistic choices~\cite{EPTCS117.7}. Further examples include process
calculi for performance evaluation, like LYSA~\cite{BODEI2005167}, proposed for
the context of cryptographic protocols, CARMA~\cite{EPTCS194.2}, specifically
defined for collective adaptive systems, MELA~\cite{EPTCS227.6}, for modeling
in ecology with location attributes, and PEPA Queues~\cite{ARGENTKATWALA20073},
introduced for the modeling of queueing networks with mobility features.
Moreover, PADS~\cite{PHILIPPOU20122} is a process algebraic framework, inspired
by real-time process algebra, for reasoning compositionally in a
component-based fashion about resource demand and supply.

A specific research topic that is worth mentioning separately is concerned with
the formal description of the behaviour of biological systems. Since the first
seminal works on the modeling and analysis of biological complex systems
through process algebra~\cite{DBLP:journals/tcs/RegevPSCS04}, several amenable
extensions of process calculi have been investigated with the aim of faithfully
specifying biological quantitative phenomena. In most cases, they are defined
in a stochastic framework supporting numerical analysis and classical
simulation techniques~\cite{doi:10.1021/j100540a008}. These studies include
stochastic versions of BioAmbients for the modeling and simulation of
biochemical reactions~\cite{VANBAKEL2008181} and of Beta-binders for the
modeling and quantitative analysis of biological systems~\cite{DEGANO2006101},
another variant of BioAmbients with context-dependent rates and ambient
volumes~\cite{BORTOLUSSI2009187}, and the stochastic calculus of wrapped
compartments~\cite{EPTCS28.6}.

To facilitate performance modeling of systems by avoiding the technicalities of
process algebra, both abstraction methods (e.g., from C source code to PEPA
models~\cite{SMITH2007129}) and mappings between high-level formalisms (e.g.,
from algorithmic skeletons to PEPA models~\cite{YAIKHOM2007167}, and from the
\textsf{nano}$\kappa$ formalism for the modeling of biochemical systems to the
Stochastic Pi Machine, a simulator for the stochastic $\pi$
calculus~\cite{LANEVE2009167}) may help the validation task performed by the
software designer/programmer. Alternative languages are proposed as a bridge
between light-weight and rigorous formalisms, like the rewrite-based specification
language PMAUDE~\cite{AGHA2006213}, encompassing both formal basis relying on
probabilistic rewrite theories and characteristics of high-level programming
languages, and supporting the specification of probabilistic concurrent
systems, discrete-event simulation, and statistical analysis.

The relations among languages and models from the expressiveness standpoint, as
well as the definition of unifying theories, have been widely investigated in
the purely nondeterministic setting, while they still represent ongoing work in
many quantitative scenarios.  To cite few examples taken from the QAPL
literature, the expressiveness comparison among languages, as proposed by
Shapiro by defining embeddings among concurrent programming
languages~\cite{10.1007/BFb0084811}, can be extended to introduce quantitative
estimates of the expressiveness similarities~\cite{BROGI2002207}.  This is done
through a notion of linear embedding by taking linear spaces as semantic
domains and, therefore, by employing a linear semantics which associates a
linear operator to each program, based on which it is possible to construct a
partial order over the languages (in the specific case, a family of Linda-like
languages). As another example, adding time to both membrane systems and Petri
nets with localities does not increase the expressiveness of the corresponding
untimed models, and a operational correspondence between these timed formalisms
can be established~\cite{EPTCS57.4}. An interesting general formal framework is
given by a modal specification theory for combined probabilistic timed systems,
called abstract probabilistic timed automata, which generalizes existing
formalisms~\cite{EPTCS117.5}.  Finally, with the aim of setting the base for
general frameworks behind labelled transition systems with quantitative
aspects, a unifying theory for nondeterministic processes with quantitative
aspects, based on a general GSOS specification format, is proposed with a
related notion of bisimulation, which induces labelled transition systems
according to the general model of ULTraS~\cite{MICULAN2016135}.

\subsection{Information flow analysis and equivalence checking}

Information flow analysis and the problem of checking the leakage of sensitive
data for programs received great attention, especially since the development of
automatic techniques that, in the early 90s, allowed for the systematic
verification of noninterference properties of security protocols and programs.
Methods to determine bounds on the amount of information leaked represented a
natural extension of the classical nondeterministic approaches, and the first
sophisticated, formal attempts in the setting of operational models of
computation go back to Gray~\cite{Gray91}, even if it is necessary to wait
another decade to see approaches dealing with quantitative analyses at the
level of the syntax of programming languages, thus posing the base for
automatic analysis of programs~\cite{VS99,Sabelfeld2000,PHW02,CLARK2002238}.
Most of these approaches can be categorized either as entropy-based information
theoretic or as inspired by probability theory. In particular, the seminal work
by Clark et al.~\cite{CLARK2002238} uses information theory, in Shannon's
sense, to analyse bounds on the actual leakage of confidential information and,
nowadays, it is still the most referenced paper in the history of QAPL, with
102 citations\footnote{Source: Scopus, May 2019.}.  The same approach is then
extended two years later to deal with loops and unbounded
iterations~\cite{CLARK2005149}, and more recently to improve
scalability~\cite{KLEBANOV2014124} and to trade for the relation existing
between scalability issues of large programs and exactness of the bounds of
leakage estimated~\cite{MU2009119}. The basic idea is to face the state space
explosion through probabilistic abstract semantics and then to estimate the
effects of the abstraction on the information flow measurement, by finding
related leakage upper bounds.  Analogously, the same quantitative information
flow approach is used in the setting of multi-threaded
programs~\cite{EPTCS117.3,EPTCS194.4}, where (probabilistic) scheduling
policies come into place, and to compute upper and lower bounds in the more
general setting of hyperproperties~\cite{YASUOKA2014167}.  Entropy-based
approaches are commonly used also in the setting of security protocol
verification, when moving from the classical symbolic approach of the Dolev-Yao
model to the computational model of security~\cite{MONTALTO2009143}.

The noninterference approach to information flow analysis has been subject to a
wide amount of work in the literature of quantitative aspects of systems, and
the classical quantitative extension of such an approach derives from the
analysis of probabilistic~\cite{ALDINI2005131},
timing~\cite{BARTHE200633,SIVERONI2006241,CHOTHIA200779}, or a mixture of
both~\cite{LANOTTE2006177} behaviours. The various approaches differ for the
modeling framework, ranging from imperative programming languages to automata
and process calculi, and for the notions of equivalence or similarity used for
analysis purposes, ranging from ad-hoc state-of-the-memory notions to
bisimulation relations.

One of the first approaches to time-sensitive noninterference for
object-oriented programming languages dealing with both exceptions and method
calls is based on an equality notion for memories and a program transformation
eliminating timing leaks~\cite{BARTHE200633}, which is still of inspiration for
several studies in time-sensitive secure information flow
analysis~\cite{Kashyap2011413} and in more practical settings, like
Java~\cite{Iranmanesh201620}, cloud and edge computing~\cite{Bazm2019197}.

In the probabilistic setting, the amount of information leakage is determined
in terms of the probability of observing a security violation and, in the case
the system execution is (partially) under the control of an adversary (which is
the typical scenario whenever the system exposes a mixture of external
nondeterministic behaviours and probabilistic behaviours), the goal becomes to
estimate the maximum leakage for the most powerful adversary. Such an objective
can be achieved, for instance, in the framework of probabilistic formal
paradigms, like process algebra~\cite{ALDINI2005131}, and exploiting
quantitative notions of equivalences, either exact or approximate.

The noninterference-based security analysis can be viewed as an instance of a
more general problem related to the quantitative comparison of the similarities
among systems. In fact, in the specific setting of information flow, the models
under comparison represent different versions of the same system with and
without the interference of events, behaviours, and/or agents that may cause the
undesired leakage. In general, the study of equivalence relations and of
alternative notions approximating the exactness conditions of such relations
played a fundamental role, and it is still mainstream, in the field of
quantitative analysis.

As far as quantitative notions of equivalence relations are concerned, in the
last 20 years several different semantics have been considered, as also
witnessed in the context of QAPL, which in such a sense proposes interesting
representatives of the lines of research in this field.  Such studies include
linear-time equivalences, like trace equivalence for continuous-time Markov
chains~\cite{WOLF2006259} and for interactive Markov chains~\cite{WOLF2006187};
probabilistic barbed congruence~\cite{DENG2007185}, which coincides with
observational equivalence for a version of CCS including a probabilistic
guarded choice operator, branching bisimulation congruence for probabilistic
transition systems obeying a general alternating model of probabilistic and
nondeterministic states~\cite{ANDOVA201258} and for a more general
probabilistic transition system specification format~\cite{EPTCS194.6}, weak
bisimulation for continuous-time Markov chains~\cite{EPTCS85.9} and for Markov
automata~\cite{EPTCS194.1}; testing equivalence for reactive probabilistic
processes~\cite{EPTCS28.7} and for nondeterministic, probabilistic, and
Markovian processes~\cite{BERNARDO20093}, reward-based testing preorders for
probabilistic labeled transition systems~\cite{DENG201416}; finally, a spectrum
of different probabilistic equivalences, including trace, bisimulation, and
testing semantics, in the setting of nondeterministic and probabilistic
processes~\cite{EPTCS117.6}, a generalized notion of bisimulation for
state-to-function transition systems that is comparable to many other
quantitative notions of bisimulation~\cite{EPTCS194.5}, and undecidability
results of bisimulation on Petri nets under durational
semantics~\cite{DBLP:journals/tcs/LasotaP16}.

When dealing with quantities, a natural extension of equivalence checking is to
relax the exactness condition based on which all the equivalence relations are
defined and to consider similarities among processes differing for negligible
details. In this setting, a typical approach borrowed from pure mathematics
relies on the use of metrics, or pseudo-metrics (see, e.g., \cite{DENG09} for a
survey related to the Kantorovich metric), which provide a measure of the
distance between non-equivalent processes.  In particular, starting from the
first developments~\cite{DBLP:conf/ifip2/GiacaloneJS90}, several definitions
converging to bisimilarity have been proposed for various models encompassing
in different ways probabilities and
nondeterminism~\cite{DENG05,ZHANG08,EPTCS194.7}, sometimes enriched with
characterizing logics~\cite{EPTCS57.11,EPTCS227.4}, while others rely on
alternative semantics, as in the case of linear/branching
distances~\cite{EPTCS57.10,Aldini201273,FAHRENBERG201454,EPTCS250.4}, as well
as game-based simulation preorders~\cite{CERNY201221}.

\subsection{Logics, with time and probability}

The need for assessing quantitative characteristics of programs and systems has
fostered not only the development of formal modeling methodologies for
behavioural specification, but also the definition of companion modeling
notations for property specification.

Several studies concern extensions of classical linear-time and branching-time
logics. In the case of Computation Tree Logic (CTL), we mention relational
abstraction techniques~\cite{HUTH200561} for hybrid CTL and hybrid Kripke
structures\footnote{We recall that hybrid approaches enrich temporal logics and
their models with the ability to name and track model states.}, and for the
probabilistic extension of the logic, PCTL, and infinite-state labelled Markov
chains~\cite{DBLP:journals/tcs/Huth05}. We then have exogenous probabilistic
CTL~\cite{BALTAZAR200795}, which is characterized by an exogenous semantics in
which the models of state formulas are probability distributions of models of a
propositional logic.  Since probabilistic CTL is not expressive enough to
reason about probabilities of sequences of formulas (like, e.g., stating that
the probability that $p$ will hold for some time is greater than the
probability that $q$ will hold for the same amount of time), more expressive
branching probabilistic logics have been proposed, like
PPL~\cite{TZANIS200879}, which is also enriched with a sound and complete
axiomatization. 

In the case of Linear Temporal Logic (LTL), one possible quantitative extension
is equipped with verification algorithms over quantitative versions of Kripke
structures and of Markov chains~\cite{FAELLA200861}. 

The $\mu$-calculus represents another important framework for quantitative
extensions, and in~\cite{MCIVER2006195} it is shown that stochastic games can
be defined in the setting of the quantitative version of the logic, with
applications to the economic domain. In order to generalize the $\mu$-calculus
and CTL to consider probabilistic aspects, it is also possible to define
extensions over
constraint-semirings~\cite{DBLP:journals/tcs/Lluch-LafuenteM05}.  The
evaluation of logic formulas is then given by values of constraint-semirings,
thus enabling in the same formalism the analysis of qualitative and
quantitative aspects. Such an approach can be applied to model quantitative
spatial aspects~\cite{CIANCIA200743}, security conditions~\cite{EPTCS194.7},
and Quality of Service (QoS) properties~\cite{HIRSCH2006135}. QoS, together
with mobility and distribution awareness, are the main subject of the logic
MoSL~\cite{DENICOLA2006161}, a temporal logic for the stochastic variant of the
process algebra KLAIM, a large fragment of which can be translated to the
Continuous Stochastic Logic (CSL).

Usability represents a fundamental issue also in the setting of property
specification, as often the gap between the practitioner and the model checking
tool prevents the former from using the latter successfully. Solutions in the
quantitative context include pattern systems~\cite{GRUHN2006117} for real-time
properties expressed in the formalism of timed automata, and a logic for the
component-oriented specification of reward-based stochastic-time
measures~\cite{DBLP:journals/tcs/AldiniB07}, that builds on a simple
first-order logic by means of which rewards are attached to the states and the
transitions of the continuous-time Markov chains (CTMCs) underlying
component-oriented system models.

\subsection{Analysis techniques and tools: program analysis, model checking, et al.}

Formal verification techniques relying on precise mathematical models provide
the base for automated analysis, which is a fundamental achievement to bridge
the gap from theory to real-world applications. 

The goal of program analysis is to establish the properties of a program
without executing it. One classical approach to semantics-based program
analysis is through abstract interpretation, which has been extended in several
different ways, including to take into account probabilistic behaviours. The
probabilistic extension leads to a semantics where programs are modeled as
linear operators represented by stochastic
matrices~\cite{DBLP:conf/ppdp/PierroW00} and, as shown for the
probabilistic $\lambda$-calculus, turns out to extend in a natural way the
classical framework~\cite{HANKIN20055}. Among the applications of such an
approach we mention the verification of resource consumption properties in the
setting of semirings~\cite{SOTIN2006153}, data flow
analysis~\cite{DIPIERRO200759} and precision analysis~\cite{DIPIERRO200823} in
the setting of the probabilistic While imperative language. An analogous approach
very close in principle is adopted to analyze quantitative properties with regard
to the use of resources such as time and memory~\cite{EPTCS28.5}.
An alternative approach, aiming to improve scalability, is based on the
abstract interpretation of a probabilistic automaton semantics for a simple
imperative language~\cite{SMITH200843}.
Alternative (numerical) program analysis techniques include robustness analysis
of program output with regard to input variations~\cite{EPTCS85.5},
probabilistic output analysis based on the input
distributions~\cite{EPTCS194.8}, lifting analysis of resource consumption from
compiled to source code~\cite{EPTCS117.2}, timing analysis of programs with
real-time constraints~\cite{LERMER20053}, and cost analysis of object-oriented
bytecode programs~\cite{ALBERT2012142}, supported by the software tool COSTA.

In contrast to static analysis, model checking is intended to verify whether a
program satisfies a property by exploring in a systematic way the state space
associated with the program. Quantitative model checking is concerned with
quantities, in most cases probabilities. The resulting algorithmic approach is
accompanied by several automated software tools, among which the probabilistic
model checker PRISM~\cite{KWIATKOWSKA20065} is one of the representatives used
in a wide range of application domains, including
security~\cite{Lanotte2005113}, safety~\cite{EPTCS28.8} and reliability of
communication protocols~\cite{Zhang2006205}. Moreover, PRISM has been extended
to deal with large, complex systems, by using, e.g., game-based
abstractions~\cite{Kattenbelt20085} and efficient algorithms based on linear
programming for the analysis of Markov decision processes
(MDPs)~\cite{GIRO201470}. 

In general, facing the state space explosion problem is an issue addressed in
the context of both purely functional and quantitative systems, and with
respect to methods like model checking of linear time and branching time
properties. Among the proposed techniques, we mention confluence reductions and
partial order reductions for branching
semantics~\cite{Baier200697,HANSEN2014103,TIMMER2016193}, and algorithms to
compute the progress of linear time model checking~\cite{EPTCS85.3}.
Alternative abstraction methods are proposed in the setting of timed
automata~\cite{EPTCS28.2} and implemented for extensions of the real-time
verification tool UPPAAL. In this framework, statistical model
checking~\cite{EPTCS85.1} is another approach used to trade between testing and
formal verification. 

In the last years, scalability of analysis has become even more relevant
because of the application of tool-supported, formal verification techniques to
domains characterized by big data evolving in time and space, like
computational biology and collective adaptive systems~\cite{EPTCS250.6}.
A widely used formalism is that of Markov population models (MPMs), for which
ad-hoc algorithms for computing transient distributions~\cite{EPTCS57.1} and to
model check CSL properties~\cite{EPTCS154.7,EPTCS194.3} have been implemented.
Enhancing scalability is the main improvement of alternative methods of
performance analysis for very large systems, e.g., based on fluid approximation
techniques~\cite{HAYDEN2012106,TSCHAIKOWSKI2014140} and on solving systems of
ordinary differential equations (ODEs)~\cite{BORTOLUSSI200727}, which is an
approach allowing for producing approximated transient measures for models of
$10^{100}$ states and beyond, even thanks to (approximate)
reductions~\cite{EPTCS250.2} and aggregations~\cite{EPTCS154.3} of ODE systems.
These methods are supported by automated analysers, like, e.g.,
GPA~\cite{EPTCS28.11}, the specification language of which is inspired by a
version of the process algebra PEPA, and
ERODE~\cite{10.1007/978-3-662-54580-5_19,EPTCS250.2}.  Approximation
methods~\cite{EPTCS57.8} and scalable reachability analysis
techniques~\cite{EPTCS250.1} are proposed also for mixed models like
probabilistic hybrid automata.

In addition to efficiency, usability is another important issue of model
checking, especially with respect to the problem of interpreting the reasons
for unsatisfiability, which has been faced in the setting of the temporal
resolution prover TRP++ for linear time properties~\cite{SCHUPPAN2016155}.

Other interesting extensions of model checking are concerned with both analysis
and modeling issues. For instance, in the former case it is worth mentioning
the analysis of infinite-state models (see, e.g., the case of CTMCs and
CSL~\cite{REMKE200724}, and the local abstraction refinement approach
developed for MDPs and reachability properties, and even implemented in 
PRISM~\cite{DRAGER201437}) and of distributed probabilistic input/output
automata~\cite{DBLP:journals/tcs/GiroDF14}, while in the latter case, some
developments emerged in the last decade are given by the analysis of
security~\cite{ADAO20063}, mobility~\cite{DENICOLA200742}, and of temporal and
epistemic properties of quantum systems, the automated verification of which is
proposed in the setting of the symbolic model checker MCMAS~\cite{EPTCS85.4}. 

Finally, studies on quantitative game theory and related algorithms for
determining the optimal
strategies~\cite{BIANCO2012160,EPTCS28.10,EPTCS57.3,EPTCS117.8,EPTCS117.9,EPTCS154.4,EPTCS227.1},
conducted for various formalisms, like, e.g., automata, are particularly
relevant, as such games can be used as models to define the interaction between
a system and its environment on the base of quantitative objectives and
behaviours inducing costs, rewards, and resource consumption. The relation with
model checking is strict, as also demonstrated by the most recent advances in
tools like, e.g., PRISM and MCMAS, implementing model checking algorithms for
MDPs founded on game semantics.

Among the other analysis techniques, supported by automated tools, we mention 
the case of language equivalence for weighted automata and of bisimulation for
conditional transition systems, both of which are accompanied with decision
algorithms implemented in PAWS~\cite{EPTCS250.5}.

\section{Conclusion, challenges, and open issues}

The field of quantitative modeling and analysis is still very active both on
the foundational aspects and on the side of software tool development.
As a representative example concerning analysis techniques, a large amount of
work has been done about notions of quantitative bisimilarity for processes
encompassing concurrency and nondeterminism, and robust extensions have been
devised in the form of behavioural pseudometrics, the kernel of which is a
bisimilarity, while in all other cases they assign a distance measuring the
degree to which two states are \textit{quasi} bisimilar.  Such metrics are
defined in terms of several alternative formulations equipped with different
(in terms of complexity) algorithms to estimate the distances.  From the
applicability standpoint, it would be useful to have a comprehensive,
comparative analysis with the aim of establishing which notions can be really
advantageous in practice. On the theoretical side, the picture of the relations
among metrics relying on different theories is not complete yet, for instance
from the viewpoint of the logical characterizations, possibly accompanied by
characteristic formulae, for such metrics. Moreover, the same effort has not
been done yet in the setting of semantics that are alternative to bisimulation,
like, e.g., testing.

Static information flow analysis is effective provided that all code is
available, while dynamic analysis considers concrete executions rather than
abstract code, even if tracking data throughout the execution process may be
difficult and time-consuming. As far as dynamic, quantitative analysis is
concerned, a large amount of modeling languages and mathematical formalisms are
nowadays accompanied by efficient verification algorithms and, therefore,
automated tools supporting their use for real-world case studies. A sign of the
reached maturity in this setting is given by the first competition among
quantitative verification tools, QComp~\cite{10.1007/978-3-030-17502-3_5},
promoted and presented in the context of TACAS during the same edition of ETAPS
in which the last QAPL has taken place. The outcome emphasizes the versatility
and the expressiveness of the considered tools, which cover an increasing
amount of specification languages, mathematical formalisms, quantitative
properties, and verification techniques.  We think that such a kind of
initiatives favours the adoption of specification standards, e.g., by promoting
the use of the JANI model exchange format, and performance standards, in order
to establish reasonable and expected tradeoffs among expressiveness, tool
efficiency, and accuracy of the results obtained.
 
Finally, it is worth observing that many of the verification techniques
discussed in this survey are close and related to other domains, as it is the
case, e.g., for probabilistic model checking, with respect to, e.g.,
probabilistic programming and (Bayesian approaches to) machine learning.  Thus,
cross-fertilisation of knowledge among the related communities may lead to
promising results. \\

\textbf{Acknowledgements:} It is worth expressing sincere thanks to all the
people who organized and put considerable effort into QAPL over the years and, 
in particular: Alessandra, Herbert, Antonio, Alessandro, Frank, Christel, Gethin, 
Mieke, Luca, Nathalie, Mirco, and Erik. 

\bibliographystyle{eptcs}
\bibliography{paper}

\end{document}